\documentstyle[12pt]{article}
\input{prepictex.tex}
\input{pictex.tex}
\input{postpictex.tex}
 \if@twoside
 \else
 \fi

 \parskip \medskipamount

\def\beq{\begin{equation}}
\def\eeq{\end{equation}}
\def\bqq{\begin{eqnarray}}
\def\eqq{\end{eqnarray}}
\def\rn{\rangle}
\def\ln{\langle}
\def\G{\Gamma}
\def\L{\lambda}

\def\bfx{\hat{{\bf x}}}
\def\bfp{\hat{{\bf \Pi}}}
\def\bfA{{\bf A}}

\def\H3{\frac{1}{H_0^3}}
\def\H2{\frac{1}{H_0^2}}

\def\bra{\langle 0|}
\def\ket{|0\rangle}
\def\brax{\langle x|}

\def\Ps{\Pi \!\!\!\! /}
\def\Ds{D \!\!\!\! /}
\def\Os{O \!\!\!\! /}
\def\gm{\gamma_{\mu } }
\def\gn{\gamma_{\nu }}
\def\ga{\gamma_{\alpha }}
\def\gb{\gamma_{\beta }}
\def\gr{\gamma_{\rho }}
\def\gd{\gamma_{\delta } }
\def\gs{\gamma_{\sigma } }
\def\hx{\hat{x} }
\def\hp{\hat{p} }
\def\psib{\bar{\psi}}

\def\dag{^{\dagger}}
\def\hX{\hat{X} }
\def\hY{\hat{Y} }
\begin{document}
\hfill{UMD-PP-95-43}
\vspace{24pt}
\begin{center}
{\bf \Large  The Inhomogeneity Expansion for  Planar Q.E.D in a Magnetic
Background}
\\
\baselineskip=12pt
\vspace{35pt}

  Gil Gat$^{1}$, Alpan Raval$^{2}$ and Rashmi Ray$^3$
\vspace{24pt}

Physics Department\\
University of Maryland\\
College Park, MD 20742\\
U.S.A\\
\vspace{130pt}
\end{center}
\begin{abstract}
The effective action for Q.E.D in external magnetic field is constructed using
 the method of inhomogeneity expansion. We first treat the non-relativistic
 case where a Chern-Simons like term is generated. We then consider the full
 relativistic theory and derive the effective action for the $A_{\mu}$ fields.
 In the non-relativistic case we also add a 4-fermi type interaction and show
that under certain circumstances, it corresponds to a Zeeman type term in the
 effective action.
\end{abstract}
\vspace{50pt}
$^1$ggat@delphi.umd.edu \\
$^2$raval@umdhep.umd.edu
\newline
$^3$rray@delphi.umd.edu

\vfill
\baselineskip=20pt
\pagebreak

1. {\bf Introduction}

The behavior of a system of interacting planar fermions in an external
magnetic field has many interesting features. In 2+1 dimensions such systems
are of relevance for the description of the Quantum Hall Effect and high
$T_{c}$ superconductivity. In addition, we have the ulterior motive
of adapting our technique to 3+1 dimensional situations. It is of
particular (and potentially of practical)
interest to study Abelian (or non-Abelian) theories involving fermions (in
3+1 dimensions) in the environment of strong magnetic fields. This is
especially rewarding in a cosmological context, where strong magnetic
fields may be present in the early universe, collapsing supernovae and
perhaps superconducting strings. To give an idea of the magnitudes of
the magnetic fields involved, we may say that those associated with
compact astrophysical objects range between $O(10^4)$ T (magnetic white
dwarfs) and $O(10^{10})$ T (supernovae)\cite {Shapiro}. In between, neutron
stars have associated magnetic fields of $O(10^8)$ T. It has also been
suggested in some models for extragalactic gamma bursts that very strong
magnetic fields $O(10^{13})$ T are involved \cite{Narayan}. The vacuum
 structure
of gauge theories in a strong magnetic background, has been and continues
to be a field of considerable interest \cite {Ambjorn},
\cite {Skal}. The minimal standard model also exhibits interesting
features in the presence of a strong magnetic background, through the
self-energy corrections of the neutrinos, through vacuum polarization
and also through non-perturbative fermion number violation \cite {Feldman},
\cite {Damgaard}. It has also been observed \cite {Krive},\cite {Klimenko}
that in certain planar fermionic theories with four-fermi interactions, a
strong magnetic field, in conjunction with a finite density, induces
breaking of the chiral symmetry and thereby generating dynamical fermion mass.
We can address all these issues through minimal modifications of the
techniques described below. The effective action in a magnetic background
plays a pivotal role in all of them.  This is one of our primary motivations in
outlining our method in considerable detail.

Effective actions neatly encapsulate
the contributions of quantum corrections to the classical action and thus
permit a classical analysis of fully quantum problems, an advantage not to
be lightly
regarded. Given an effective action for a quantum theory, one can read off
the various physical quantities of interest like effective masses, permeability
and susceptibility with relative ease. In this paper, we wish
to find the effective action of the system in terms of the fluctuating gauge
fields that the planar fermions are subjected to, by tracing the fermions
out of the partition function. The formal process of integrating Grassman
valued fields out yields a functional determinant, generically a nonlocal
quantity.
In order to obtain a Lagrangian comprising local fields and their derivatives,
we have to expand the determinant within some controlled scheme. The
derivative expansion method, also called the inhomogeneity expansion in the
present context, is one such scheme.
To implement this, we assume that the
fluctuating fields coupled to the fermions vary slowly in space-time compared
to the natural scales set by the background magnetic field. Namely, the typical
 momenta associated with these fluctuations are $ \ll \frac{1}{l}$, where
$l \equiv \frac{1}{\sqrt{eB}}$ is the magnetic length and the typical
 frequency
is $\omega \ll \omega_{c}$, where $\omega_{c} \def \frac{eB}{m}$ is the
cyclotron frequency. This naturally sets a limit on contributions from the
higher derivatives of the electromagnetic fields by filtering out and
discarding the high frequency and the high momentum modes and thus enables
us to write a local Lagrangian containing the desired number of derivatives
of the field.

 Our method obviously works well in the absence
of gaps in the spectrum and fails precisely at the points where such gaps
exist. Particularly, in this problem, where the spectrum is described by
Landau levels, the method expectedly fails  where the gap between
successive levels is probed. In practical terms, we see, in our subsequent
calculations, the emergence of spurious singularities in the form of
derivatives
of the delta function precisely at these points. In this work, we have adopted
the somewhat pragmatic stance of disregarding these singularities and tacitly
admitting the shortcomings of our method at these particular points. We however
hasten to add that away from these particular points, derivative expansion
is a perfectly reliable method and leads to a rather formulation of the
effective action. If thermal effects are taken into account resulting in
the smearing out of these discrete levels through thermal fluctuations,
this method can be readily applied to even the points mentioned earlier.

Although the solution of the ``free'' single particle problem is known in this
case, the construction of an effective action (or potential) for the
interacting  theory is rather involved. In particular since the theory is
gauge invariant to begin with, one would like to find a method or rather an
expansion, that will exhibit this gauge invariance order by order in a
manifest way.

 In this paper we construct such a method. We apply it to the nonrelativistic
 test case\footnote{This problem was originally considered by Ray
and Sakita \cite{Pani} for the non zero temperature case. Our results agree
 with
theirs for the T=0 case (see also \cite{saki})} before  tackling the fully
relativistic
theory.
The basic idea of our method is the use
of guiding center coordinates which are the natural variables for this
problem. Using these variables, we are able to convert the problem of
calculating
the fermion determinant into the problem of calculating matrix elements
of known commutators and of lowering and raising operators. In addition we make
several unitary transformations along the way to expose the inherent gauge
invariance of the system  at each stage of the approximation.

The organization of this paper is as follows: we start with the nonrelativistic
 problem, establishing the notation and describing
the approximations entailed and transformations made in section 2. In section
3 we present the calculations and the results for the nonrelativistic case.
In section 4 we study the effect of adding a four Fermi type
interaction to the original Lagrangian and show that in a nonrelativistic
theory this kind of term, under certain circumstances, can induce a Zeeman
type interaction in the effective Lagrangian. In section 5 we
consider the fully relativistic QED .
 We conclude with
discussions in section 6. Useful relations are deferred to the appendices.

2. { \bf Nonrelativistic Q.E.D}

To set the stage for the calculation we start by establishing the notation.
Consider the single particle Landau problem.
The dynamics of a single planar particle in an uniform magnetic field normal
to the plane is described by the Landau hamiltonian:
\[ H_0=\frac{1}{2m} {\bf P}^2 \]
where ${\bf A}$ may be chosen in a specific gauge ${\bf A}=B(-y,x)$. The
wave functions of  the system, $ \brax n,X \rn$, are labeled by two parameters;
$n$ and $X$. $n$, which is an integer is the Landau level index and $X$, a
continuous parameter, measures the degeneracy of any given Landau level.
The single particle energies are given by:
\beq
E_n=(n+\frac{1}{2})w_c
\eeq
where $w_c=\frac{eB}{m}$ is the cyclotron frequency.
These energies are obviously independent of $X$ and the operator $\hat{X}$
which measures the value of $X$ commutes with the hamiltonian. We call
$\hat{X}$ and its conjugate $\hat{Y}$ the guiding center coordinates.They
play a central role in our subsequent discussions.

Let us define $\bfp_1=\hat{p}_1+\frac{eb}{2}\bfx_2$ and $\bfp_2=\hat{p}_2
-\frac{eb}{2}\bfx_1$ where $[\bfp_1, \bfp_2]=\frac{i}{l^2}$.
We can similarly define $\hat{a}=\frac{l}{\sqrt{2}}(\bfp_1+i\bfp_2)$.
The guiding center operators are then defined as
 \beq \hat{X}=\bfx_1+l^2 \bfp_2 \hspace{2cm}\mbox{and}\hspace{2cm}
\hat{Y}=\bfx_2-l^2 \bfp_1 \label{xydef}
\eeq
thus  $ [\hat{X},\hat{Y}]=-il^2  $. These are the two sets of cannonically
conjugate operators that we need:
\[
\hat{a}\dag \hat{a} |n,X \rn =n|n,X \rn \hspace{3cm} \hat{X}|n,X \rn =
X|n,X \rn
\]
The normalization for $X$ is fixed by the degeneracy of the Landau levels
per unit area. Namely
\beq
\int dX |\ln n,X|n,X \rn |^2=\frac{1}{2 \pi l^2}
\eeq
where $\brax n,X \rn = \frac{1}{\sqrt{2 \pi l^3}}e^{\frac{iXy}{l^2}}u_n (\frac{
x-X}{l})$ and $u_n(\frac{x-X}{l})$ are the standard harmonic oscilator wave
functions.

Using an euclidean metric, the partition function for the corresponding
many particle system is given by:
\beq
{\cal Z}=\int D\psi D \psi\dag e^{-\int d\vec{x}dt \psi\dag \{
\Pi_{\tau}+\frac{1}{2m}({\bf \Pi})^2-\mu \} \psi }
\eeq
where $\hat{\Pi}_{\tau}=-i\partial_{\tau}-iA_{0}=
\hat{p}_{\tau}-eA_{\tau}$ ,and
 $\bfp=\hat{{\bf p}}_{\tau} -e{\bf A}$.
Integrating the fermions out we get ${\cal Z}=e^{-S_{eff}}$ where
\beq
S_{eff}=-\mbox{Tr} \log [i \hat{\Pi}_{\tau}+\frac{1}{2m} \bfp^2-\mu ]
\eeq
The corresponding current is
\[
\langle J_{\mu}(x)\rangle \equiv \frac{\delta S_{eff}}{\delta A_{\mu}(x)}
\]
where $J_{\mu}(x)=(ie \frac{1}{i\hat{\Pi}_{\tau}+\frac{1}{2m}\bfp^2-\mu}
,\{ \bfp_i,\frac{1}{i\hat{\Pi}_{\tau}+\frac{1}{2m}\bfp^2-\mu}  \}) $.
We translate the bra and ket to $x=0$, consequently the operators
 $\hat{x}_{\mu}$ in $J_{\mu}$ get translated to $\hat{x}_{\mu}+x_{\mu}$
 and we expand all the functions of $\hat{x}_{\mu}+x_{\mu}$ around
$x_{\mu}$ (derivative expansion).

To make the gauge invariance as manifest as possible perform a succession  of
 unitary transformations $\Gamma=WVU$, on $J_{\mu}$ where
\bqq
U(x,t)&=&e^{-ie(\bfx \cdot  \bfA  +\frac{1}{2!} \bfx_i \bfx_j \partial_i
\partial_j
\bfA_j+\frac{1}{3!} \bfx_i \bfx_j  \bfx_k \partial_i \partial_j \partial_k \bfA
+\ldots )}
\\ \nonumber
V(x,t)&=&e^{-ie[\hat{\tau} A_{\tau}(x)+\frac{\hat{\tau}^2}{2!}
\dot{A}_{\tau}(x)+
\frac{\hat{\tau}^3}{3!}\ddot{A}_{\tau}+\ldots }
  \\ \nonumber
W(x,t)&=&e^{\frac{i}{2l^2}\bfx_1 \bfx_2}
\eqq
We notice that $\Gamma$, acting on $|x_{\mu}=0 \rn$, produces just a factor of
unity\footnote{At this point we note that the transformation $V$ fails to be
 unitary in the corresponding finite temperature problem due to the
compactness of the domain of definition of $\tau$ , $(0 \rightarrow
 \beta=1/T)$. At $T=0$, however, $V$ is unitary.}.
Thus
\beq
\ln J_{\mu}(x,t) \rn = \ln 0|\Gamma \tilde{J}_{\mu} \Gamma\dag |0\rn .
\eeq
Since $A_{\tau}$ is entirely fluctuating, we refer to it as $a_{\tau}$,
similarly the
fluctuating part of ${\bf A}_i$ is refered to as ${\bf a}_i$.

After all these transformations we obtain:
\bqq
\ln J_{\tau}\rn &=&ie \ln 0|\frac{1}{i\hat{\Pi}_{\tau}-\mu+i\Delta_{\tau}+
\frac{1}{2m}(\bfp+{\bf \Delta})^2}|0\rn  \\ \nonumber
\langle J_i \rangle &=& \frac{e}{2m} \ln 0| \left\{ \bfp_i,\frac{1}{ H_0
+\frac{1}{2m} ( \bfp \cdot
{\bf \Delta}+{\bf \Delta} \cdot \bfp +{ \bf \Delta}^2 )+i\Delta_{\tau}}
 \right\} |0\rn  \label{jays}
\eqq
where $H_0=i \Pi_{\tau}+\frac{1}{2m} \bfp^2-\mu$ ,
\[ H_0 |n,X,w \rangle =\Gamma_n |n,X,w \rangle =[iw-\mu+(n+\frac{1}{2})w_c
] |n,X ,w
\rn \] and
\bqq
  \Delta_{\tau}&=&-e[\bfx  \cdot {\bf \partial} a_{\tau}+\frac{1}{2}\bfx_i
 \bfx_j
 \partial_i \partial_j a_{\tau} +\bfx_i \hat{\tau} a_i a_{\tau}+ \ldots ]
\\ \nonumber
  {\bf \Delta}_l&=&-e[\hat{\tau} \dot{a}_l+\frac{1}{2} \hat{\tau}^2
\ddot{a}_l+\hat{\tau}  \bfx
\cdot {\bf \partial} \dot{a}_l+\frac{1}{2} \bfx_i {\cal
F}_{il}+\frac{1}{3}\bfx_i \bfx_j
{\bf \partial}_i {\cal F}_{jl}+\ldots ].
\eqq
 $H_0$ can be solved  exactly (its spectrum being the L.L) and the rest is
treated
perturbatively in the following section. In fact,
\beq
H_0 |n,X,w \rn =\Gamma_n |n,X,w \rn =[i w -\mu +(n+\frac{1}{2})w_c]|n,X,w \rn
\, \, .
\eeq

3. {\bf Effective Action }

The most efficient way of computing the effective action is to obtain
 $\ln \Pi_{\mu \nu} \rn$ the
polarization tensor, from $\ln J_{\mu} \rn$ and write:
\beq
S_{eff}=\frac{1}{2} \int dx dy \, a_{\mu}(x) \ln \Pi_{\mu \nu}(x-y) \rn
a_{\nu}(y)
\eeq
where $\ln \Pi_{\mu \nu} \rn =\frac{\delta \ln J_{\mu}(x) \rn }{\delta
a_{\nu}(y)}$. Using
 the remaining rotational symmetry, the only independent components of
 $\ln \Pi_{\mu \nu} \rn$ that are
 needed to reconstruct the effective action are $\ln \Pi_{00} \rn$,
$\ln \Pi_{01} \rn$,
 $\ln \Pi_{11} \rn$ and $\ln \Pi_{12} \rn$.
In section 2 we transformed $\ln J_{\mu} \rn$ to a form that allows the use of
 perturbation theory. Thus for example  we can expand
eq.(\ref{jays}) to get
\bqq
\ln J_{\tau} \rn
& &ie \ln 0|\frac{1}{i\hat{p}_{\tau}-\mu+i\Delta_{\tau}+
\frac{1}{2m}({\bf \hat{\Pi}}+{\bf \Delta})^2}|0 \rn  \\ \nonumber
&=&ie \ln 0|\frac{1}{H_0}|0 \rn -ie \ln 0|\frac{1}{H_0} \left( i
\Delta_{\tau}+\frac{1}{2m}
({\bf \hat{\Pi}} \cdot {\bf \Delta}+{\bf \Delta \cdot \hat{\Pi}}+{\bf
 \Delta}^2 ) \right)\frac{1}{H_0}|0 \rn +\ldots
\label{Jt}
\eqq
Now the $\Delta_{\mu}$ generically contain $\hat{x}_{\mu}$. We use the
 definitions of
$\hat{X}$ and $\hat{Y}$ from section 2 to convert the $\hat{x}_{\mu}$ into
$\hat{X}$ and $\hat{Y}$ and $\bfp_i$. The action of $\hat{X}$ and $\hat{Y}$ on
$|x_{\mu} \rn =0$ (or its complex conjugate)  enables us to get rid of
$\hat{X}$ and
$\hat{Y}$ altogether and obtain an operator that depends only on $\bfp$.
 Subsequently
we can introduce a complete set $\{ |n,X \rn \}$ of states and integrate over
$X$ (as the operator does not contain any further guiding center coordinates)
immediately to obtain the degeneracy factor $\rho_0=\frac{1}{2\pi l^2}$.
The remaining evaluation of harmonic oscillator matrix elements is
straightforward.

To clarify the foregoing statements through an example, we compute the
contribution
proportional to $f_{12}$ to $\ln J_{\tau} \rn$.
This is given by
\[ \frac{ie^2}{2m} \ln 0| \frac{1}{H_0} (\bfx_1 \bfp_2-\bfx_2 \bfp_1)
\frac{1}{H_0}|0\rn \]
now, $\bfx_1 =\hat{X}-l^2 \bfp_2$ and $\bfx_2=\hat{Y}+l^2\bfp_1$ , therefore
\[ \ln 0|\hat{X}=l^2 \ln 0|\bfp_2  \hspace{3cm} \ln 0| \hat{Y} =-l^2
 \ln 0|\bfp_2 \,\, . \]
 Also
$[\bfp_1,\frac{1}{H_0}]=-\frac{i}{ml^2} \frac{1}{H_0} \bfp_2 \frac{1}{H_0}$
 and similarly for $\bfp_2$, $[\bfp_2,\frac{1}{H_0}]=\frac{i}{ml^2}
\frac{1}{H_0} \bfp_1 \frac{1}{H_0}$. Thus the above matrix element becomes:
\bqq
& &\frac{ie^2l^2}{2m} \ln 0|[\bfp_2,\frac{1}{H_0}]\bfp_2 \frac{1}{H_0}
+[\bfp_1,
\frac{1}{H_0}]\bfp_1 \frac{1}{H_0} |0 \rn \\ \nonumber
&=& -\frac{e^2}{2m^2} \ln 0|\frac{1}{H_0}\bfp_1\frac{1}{H_0}\bfp_2
\frac{1}{H_0}
- \frac{1}{H_0}\bfp_2  \frac{1}{H_0}\bfp_1 \frac{1}{H_0}|0 \rn   \\ \nonumber
&=& -\frac{e^2}{m^2} \ln 0|\frac{1}{H_0}\bfp_1\frac{1}{H_0}\bfp_2
\frac{1}{H_0}|0 \rn  \\ \nonumber
&=& -\frac{e^2}{m^2} \sum_{n_1,n_2,n_3} \int dX \int \frac{dw}{2 \pi}
\frac{ \ln 0|n_1,X,w \rn \ln n_3,X,w|0 \rn }{\Gamma_{n_1} \Gamma_{n_2}
\Gamma_{n_3}} \ln n_1|
\bfp_1 |n_2 \rn \ln n_2|\bfp_2 |n_3 \rn
\eqq
The integration over $X$ can now be done trivially yielding $\frac{1}{2 \pi
l^2} \delta_{n_1,n_3}$, reducing the matrix element to:
\[ -\frac{e^2}{2 \pi m^2 l^2} \sum_{n_1,n_2} \int \frac{dw}{2 \pi} \frac{
\ln n_1|\bfp_1 |n_2 \rn \ln n_2|\bfp_2 |n_3 \rn }{\Gamma_{n_1}^2 \Gamma_{n_2} }
\]
Using the definition of $\bfp$ we get
\beq
-\frac{ie^2}{4 \pi m^2 l^4} \int \frac{dw}{2 \pi} \sum_{n=0}^{\infty} (n+1)
\left[ \frac{1}{\Gamma_n^2 \Gamma_{n+1}}-\frac{1}{\Gamma_n \Gamma_{n+1}^2}
\right]
\eeq
and we use $\frac{1}{\Gamma_n \Gamma_{n+1}}=\frac{1}{w_c} \left[
\frac{1}{\Gamma_n}-\frac{1}{\Gamma_{n+1}} \right]$
Thus the above matrix element reduces to (retaining only the non-singular
part as discussed in the introduction)
\[
\frac{ie^2}{2 \pi} \sum_{n=0}^{\infty} \int \frac{dw}{ 2 \pi} \frac{1}{iw-\mu
+(n+\frac{1}{2})w_c} f_{12}(\bfx, \tau)
=\frac{ie^2}{2 \pi}\sum_{n=0}^{\infty}
\Theta (w_c (n+\frac{1}{2})-\mu) f_{12}(\bfx, \tau)
\]
For operators involving $\hat{\tau}$ we use $[\hat{\tau}, \frac{1}{H_0}]=
\frac{1}{H_0^2}$ and proceed as above.
Another simplifying feature is that any matrix element with an odd number of
$\bfp_i$ is zero.

It is easy to show that all terms which could possibly lead to  gauge
noninvariance vanish identically. We append an example here.
The term proportional to $\ddot{a}_i$ in
$\langle J_{\tau} \rangle $ leads to a  gauge
noninvariant term in the effective action. The matrix element proportional to
this term is
\beq
\frac{ie^2}{2m} \ln 0| \frac{1}{H_0} \bfp_1 \hat{\tau}^2   \frac{1}{H_0} |0 \rn
\ddot{a}_1+ (1\rightarrow 2)
\eeq
Now since $\hat{\tau}^2 \frac{1}{H_0}|0\rn=\frac{2}{H_0^3}|0 \rn$ we get
\beq
\frac{ie^2}{m}\ln 0| \frac{1}{H_0} (\bfp_1 \ddot{a}_1+\bfp_2 \ddot{a}_2 )
 \frac{1}{H_0}|0 \rn =0
\eeq
so that in fact this term does not appear.

At this point we collect all the relevant terms, i.e. those proportional to
$\partial_1^2 a_{\tau}$, $\partial_1 \partial_{\tau} a_{1}$ and $f_{12}$ for
$\langle J_{\tau} \rangle$ and proportional to $\partial_{\tau}^2 a_{1}$,
$\partial_{\tau}^2
 a_{2}$, $\partial_1 \partial_{\tau} a_{\tau}$, $\partial_2 \partial_{\tau}
 a_{\tau}$ and $\partial_2 f_{12}$ for $ \ln J_1 \rn$ and obtain :
\bqq
\ln \Pi_{\tau \tau}(x,y) \rn &=& \frac{e^2}{2 \pi w_c}C (\partial_1^2+
\partial_2^2)\delta (x-y)+\ldots
\\ \nonumber
\ln \Pi_{\tau 1}(x,y)\rn  &=& -\frac{e^2 }{2 \pi w_c} C \partial_{\tau}
\partial_1 \delta (x-y)-
\frac{ie^2}{2 \pi} C \partial_2 \delta (x-y)+\ldots
\\ \nonumber
\ln \Pi_{\tau 2}(x,y)\rn &=& -\frac{e^2}{2 \pi w_c}  C \partial_{\tau}
\partial_2+\frac{ie^2}{2 \pi}C \partial_1 \delta (x-y) +\ldots
\\ \nonumber
\ln \Pi_{11}(x,y)\rn &=& \frac{e^2}{2 \pi w_c} C \partial_{\tau}^2 \delta (x-y)
-\frac{e^2}{\pi m}C_1 \partial_2^2 \delta (x-y)+\ldots
\\ \nonumber
\ln \Pi_{12}(x,y)\rn &=&\frac{e^2}{\pi m} C_1 \partial_1 \partial_2
\delta (x-y)
-\frac{ie^2}{2 \pi} C \partial_{\tau} \delta (x-y)+\ldots
\\ \nonumber
\ln \Pi_{22}(x,y) \rn &=&\frac{e^2}{2 \pi w_c}C \partial_{\tau}^2 \delta (x-y)-
\frac{e^2}{\pi m}C_1 \partial_1^2 \delta (x-y)+\ldots
\eqq
where
\beq
C= \sum_{n=0}^{\infty} \int \frac{dw}{2 \pi} \left( 1+w_c
 \frac{\partial}{\partial w_c }\right) \frac{1}{\Gamma_n} =\sum_{n=0}^{\infty}
\Theta (\mu-(n+\frac{1}{2})w_c)+\ldots
\eeq
and $C_1$ is
\beq
C_1=\sum_{n=0}^{\infty}(n+\frac{1}{2}) \int \frac{dw}{2 \pi} \left(
 1+\frac{1}{2}w_c
 \frac{\partial}{\partial w_c }\right) \frac{1}{\Gamma_n}
 =\sum_{n=0}^{\infty}(n+\frac{1}{2})
\Theta (\mu-(n+\frac{1}{2})w_c)+\ldots
\, \, .
\eeq
The effective action is reconstructed as :
\beq
S_{eff}=\frac{ie^2}{2 \pi} C \epsilon_{\mu \nu \rho} \int d^3x a_{\mu}
 \partial_{\nu} a_{\rho} + \frac{e^2}{4 \pi w_c}C \int
 d^3x \,{\cal E}^2+\frac{e^2}{2 \pi m}C_1 \int d^3x \, {\cal B}^2+ \ldots
\eeq
where $\vec{{\cal E}}$ and ${\cal B}$ are the euclidean fluctuating electric
and
magnetic field strength respectively. If we choose  $\mu$ such that exactly $N$
Landau Levels are filled
$C \rightarrow N+1$ and $C_1 \rightarrow \frac{(n+\frac{1}{2})^2}{2}$.

We note two points regarding this effective action. The first is the
emergence of a C-S term. The second is that since
relativistic invariance is broken by the external magnetic field,
we get different coefficients for the fluctuating ${\bf {\cal E}}^2$ and
${\cal B}^2$ terms. Further if the magnetic field is very strong (i.e.
$l \ll \frac{1}{m}$), the second term in the effective action does not
contribute. This corresponds to dimensional reduction. We also observe
that if there are no
fermions ,i.e. $\mu=0$, the magnetic ${\cal B}^2$ term in
the gauge field action is not renormalized by quantum effects. This is to
be contrasted with the relativistic case (see eq.(\ref{relef})) where the
vacuum (filled
Dirac sea) does contribute to the renormalization of this term.

We can use this effective action (rotated to minkowski space) to obtain
physically interesting quantities,
for example the dielectric function
\[
\epsilon({\bf p})= 1+\frac{\Pi_{00}(0,{\bf p})}{{\bf p^2}}
=1+\frac{e^2}{2 \pi w_c}C
\]
and the magnetic permeability
\[
\frac{1}{\mu(0,{\bf p})} =1+\frac{\Pi_i^{i}(0,{\bf p})}{{\bf p^2}}=
1-\frac{e^2}{ \pi m} C_1  \, \, .
\]

4. {\bf Effective Zeeman Interaction}

Having obtained the effective action we take this oportunity to explore an
interesting phenomenon namely the effect of adding a four fermi term.
 In this section we show that if a four fermi term is added to the original
 Lagrangian introduced in the previous sections, an effective Zeeman-type
interaction could be induced in certain cases.

The partition function for the system is
\bqq
{\cal Z} &=& \int D \phi D\psi D{\psi }^{\dagger }e^{S(\phi, \psi \psib)} \\
\nonumber
S&=& i\int d^3x \left[ {\psi }^{\dagger }[i\partial_0-eA_0+\mu +
\frac{1}{2m}(i{\bf \nabla} + e{\bf A})^2]\psi
-g({\psi }\dag \frac{\sigma_3}{2}\psi )^2\right]
\eqq
Using an auxiliary field $\phi $ to disentangle the quartic term, we obtain
\beq
 {\cal Z} = \int D\psi D{\psi }^{\dagger }exp\left\{ i\int d^3x \left[
\psi^{\dagger
 }[ i\partial_0-eA_0+\mu +\frac{1}{2m}(i{\bf \nabla } + e{\bf A})^2
+\sqrt {\frac{g}{2}}\phi \sigma_3]\psi +\frac{1}{2} \phi^2 \right]  \right\}
\eeq
We wish to show that the effective potential for $\phi $ obtained upon
integrating the fermions out is minimized by a non-zero constant configuration
which we call $\phi_0 $. This in turn would mean that the Yukawa term
in the action above yields an effective Zeeman term $\sqrt{\frac{g}{2}}
\phi_0 \psi \dag \sigma_3 \psi $.

The effective potential for $\phi $ is given by
\beq
V_{eff} = -\frac{1}{2\pi l^2} \sum_{n=0}^{\infty } \int \frac{d k_0}{2 \pi i}
\,  \log \left[ \frac{k_0-E_n+\mu +\frac{b}{2}}{k_0-E_n+\mu -\frac{b}{2}}
 \right],
\eeq
where $E_n=\frac{1}{ml^2}(n+\frac{1}{2})$ and $b \equiv \sqrt{2g}\phi $.
The presence of the sum of two logarithms is due to the two component nature
of the fermion.
The mean density is computed from $V_{eff}$ as
$\bra \rho \ket =-{{\partial V}\over{\partial \mu }}$, which yields
\beq
\bra \rho \ket =\frac{1}{2\pi l^2} \sum_{n=0}^{\infty } \int \frac{d k_0}{2\pi
 i} \left[ \frac{1}{k_0-E_n+\mu +\frac{b}{2}}+\frac{1}{k_0-E_n
 +\mu -\frac{b}{2}} \right] .
\eeq

\begin{figure}
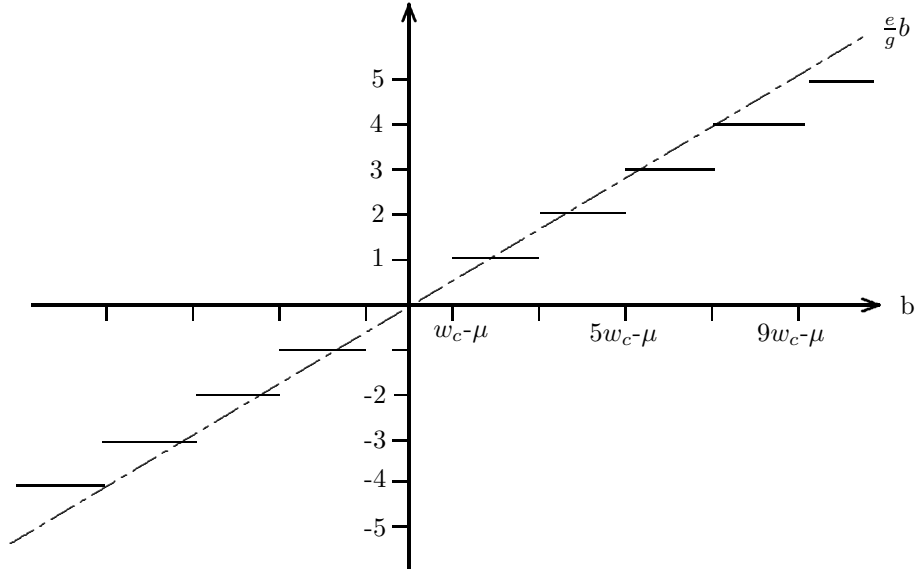

\begin{center}
\mbox{\beginpicture
\setcoordinatesystem units <.25mm,.25mm>
\unitlength=.25mm
\linethickness = .5pt
\setplotsymbol({\fiverm .})
\setplotarea x from  77  to  558 , y from 0 to  310
\typeout{\space\space\space Picture exported by 'qfig'.}
\font\FonttenBI=cmbxti10\relax
\font\FonttwlBI=cmbxti10 scaled \magstep1\relax
\setlinear
\putrule from  174  139  to  174  139
\putrule from  174  139  to  174  132
\setlinear
\putrule from  128  139  to  128  139
\putrule from  128  139  to  128  132
\setlinear
\putrule from  220  139  to  220  139
\putrule from  220  139  to  220  132
\setlinear
\putrule from  174  139  to  174  139
\putrule from  174  139  to  174  132
\setlinear
\putrule from  266  139  to  266  139
\putrule from  266  139  to  266  132
\setlinear
\putrule from  220  139  to  220  139
\putrule from  220  139  to  220  132
\setlinear
\putrule from  312  139  to  312  139
\putrule from  312  139  to  312  132
\setlinear
\putrule from  266  139  to  266  139
\putrule from  266  139  to  266  132
\setlinear
\putrule from  358  139  to  358  139
\putrule from  358  139  to  358  132
\setlinear
\putrule from  312  139  to  312  139
\putrule from  312  139  to  312  132
\setlinear
\putrule from  404  139  to  404  139
\putrule from  404  139  to  404  132
\setlinear
\putrule from  358  139  to  358  139
\putrule from  358  139  to  358  132
\setlinear
\putrule from  450  139  to  450  139
\putrule from  450  139  to  450  132
\setlinear
\putrule from  404  139  to  404  139
\putrule from  404  139  to  404  132
\setlinear
\putrule from  496  139  to  496  139
\putrule from  496  139  to  496  132
\setlinear
\putrule from  450  139  to  450  139
\putrule from  450  139  to  450  132
\linethickness = 1 pt
\setplotsymbol({\makebox(0,0)[l]{\tencirc\symbol{'160}}})
\setlinear
\putrule from  88  140  to  88  140
\putrule from  88  140  to  539  140
\linethickness = .5pt
\setplotsymbol({\fiverm .})
\linethickness = 1 pt
\setplotsymbol({\makebox(0,0)[l]{\tencirc\symbol{'160}}})
\setlinear
\putrule from  289  140  to  289  140
\putrule from  289  140  to  289  300
\linethickness = .5pt
\setplotsymbol({\fiverm .})
\linethickness = 1 pt
\setplotsymbol({\makebox(0,0)[l]{\tencirc\symbol{'160}}})
\setlinear
\putrule from  289  141  to  289  141
\putrule from  289  141  to  289  0
\linethickness = .5pt
\setplotsymbol({\fiverm .})
\linethickness = 1 pt
\setplotsymbol({\makebox(0,0)[l]{\tencirc\symbol{'160}}})
\setlinear
\plot  285.7508  291.0729  289  300 /
\plot  289  300  292.2492  291.0729 /
\linethickness = .5pt
\setplotsymbol({\fiverm .})
\setlinear
\putrule from  220  116  to  220  116
\putrule from  220  116  to  266  116
\setlinear
\putrule from  176  92  to  176  92
\putrule from  176  92  to  220  92
\setlinear
\putrule from  312  165  to  312  165
\putrule from  312  165  to  358  165
\setlinear
\putrule from  359  189  to  359  189
\putrule from  359  189  to  404  189
\setlinear
\putrule from  126  67  to  126  67
\putrule from  126  67  to  176  67
\setlinear
\putrule from  127  44  to  127  44
\putrule from  127  44  to  80  44
\setlinear
\putrule from  404  212  to  404  212
\putrule from  404  212  to  451  212
\setlinear
\putrule from  451  236  to  451  236
\putrule from  451  236  to  498  236
\setlinear
\putrule from  460  236  to  460  236
\putrule from  460  236  to  499  236
\setdashpattern < 10pt, 3pt, 2pt, 3pt >
\setlinear
\putrule from  77  13  to  77  13
\putrule from  77  13  to  77  13
\plot  77  13  531  283 /
\setsolid
\setlinear
\putrule from  280  46  to  280  46
\putrule from  280  46  to  280  46
\putrule from  280  46  to  288  46
\setlinear
\putrule from  280  68  to  280  68
\putrule from  280  68  to  280  68
\putrule from  280  68  to  288  68
\setlinear
\putrule from  281  92  to  281  92
\putrule from  281  92  to  281  92
\putrule from  281  92  to  288  92
\setlinear
\putrule from  280  116  to  280  116
\putrule from  280  116  to  280  116
\putrule from  280  116  to  288  116
\setlinear
\putrule from  280  140  to  280  140
\putrule from  280  140  to  280  140
\putrule from  280  140  to  288  140
\setlinear
\putrule from  281  164  to  281  164
\putrule from  281  164  to  281  164
\putrule from  281  164  to  288  164
\setlinear
\putrule from  280  164  to  280  164
\putrule from  280  164  to  280  164
\putrule from  280  164  to  288  164
\setlinear
\putrule from  281  188  to  281  188
\putrule from  281  188  to  281  188
\putrule from  281  188  to  288  188
\setlinear
\putrule from  280  188  to  280  188
\putrule from  280  188  to  280  188
\putrule from  280  188  to  288  188
\setlinear
\putrule from  281  212  to  281  212
\putrule from  281  212  to  281  212
\putrule from  281  212  to  288  212
\setlinear
\putrule from  280  212  to  280  212
\putrule from  280  212  to  280  212
\putrule from  280  212  to  288  212
\setlinear
\putrule from  281  236  to  281  236
\putrule from  281  236  to  281  236
\putrule from  281  236  to  288  236
\setlinear
\putrule from  280  236  to  280  236
\putrule from  280  236  to  280  236
\putrule from  280  236  to  288  236
\setlinear
\putrule from  279  236  to  279  236
\putrule from  279  236  to  279  236
\putrule from  280  236  to  288  236
\setlinear
\putrule from  280  260  to  280  260
\putrule from  280  260  to  280  260
\putrule from  280  260  to  288  260
\setlinear
\putrule from  281  22  to  281  22
\putrule from  281  22  to  281  22
\putrule from  281  22  to  289  22
\setlinear
\putrule from  280  22  to  280  22
\putrule from  280  22  to  280  22
\putrule from  280  22  to  288  22
\setlinear
\putrule from  281  46  to  281  46
\putrule from  281  46  to  281  46
\putrule from  281  46  to  289  46
\setlinear
\putrule from  280  46  to  280  46
\putrule from  280  46  to  280  46
\putrule from  280  46  to  288  46
\put{{\xpt\rm 1}}[lt] at  269  170
\put{{\xpt\rm 2}}[lt] at  269  193
\put{{\xpt\rm 3}}[lt] at  268  217
\put{{\xpt\rm b}}[lt] at  550  145
\linethickness = 1 pt
\setplotsymbol({\makebox(0,0)[l]{\tencirc\symbol{'160}}})
\setlinear
\plot  530.0729  143.2492  539  140 /
\plot  539  140  530.0729  136.7508 /
\linethickness = .5pt
\setplotsymbol({\fiverm .})
\put{{\xpt\rm $w_c$-$\mu$}}[lt] at  302  129
\put{{\xpt\rm $5w_c$-$\mu$}}[lt] at  385  130
\put{{\xpt\rm $9w_c$-$\mu$}}[lt] at  474  130
\setlinear
\putrule from  502  259  to  502  259
\putrule from  502  259  to  536  259
\put{{\xpt\rm 4}}[lt] at  268  241
\put{{\xpt\rm 5}}[lt] at  269  265
\put{{\xpt\rm -2}}[lt] at  265  97
\put{{\xpt\rm -3}}[lt] at  265  73
\put{{\xpt\rm -4}}[lt] at  265  52
\put{{\xpt\rm -5}}[lt] at  265  26
%
\put{{\xpt\rm $\frac{e}{g}b$}}[lt] at  540  293
\endpicture}
\end{center}
\caption{Graphical solution of eq.(30). The $y$ axis is labeled in units of
$\rho_0=\frac{eB}{2 \pi}$ }
\end{figure}
On performing the $k_0$ integral, we get
\beq
\bra \rho \ket \equiv \bra \rho \ket_{+} + \bra \rho \ket_{-},
\eeq
where
\beq
\bra \rho \ket_{+}\equiv \frac{1}{2\pi l^2}\sum_{n=0}^{\infty } \theta (\mu
+{b\over 2}-E_n)
\eeq
\beq
\bra \rho \ket_{-}\equiv {1\over{2\pi l^2}}\sum_{n=0}^{\infty } \theta (\mu
-{b\over 2}-E_n)
\eeq
We reconstruct $V_{eff}$ from $\bra \rho \ket $ using
\beq
V_{eff}=-\bra \rho \ket A_0 - {1\over{4g}}b^2 ,
\eeq
where the second term is the tree level contribution.
We note that in the absence of fluctuating electrmagnetic fields, the role
of $A_0$ is played by $-{1\over{2e}}b$ in $\bra \rho \ket_{+}$ and by
${1\over{2e}}b$ in $\bra \rho \ket_{-}$.
Thus
\beq
V_{eff}(b)={1\over{2e}}b \, (\bra \rho \ket_{+}-\bra \rho \ket_{-})-
{1\over{4g}}b^2 .
\eeq
We wish to look for extrema of the above away from the points
$b=\pm [(2n+1){1\over{ml^2}}-2\mu ]$, since at these points the extrema picks
up singular delta function contributions.
The extrema satisfy
\beq
\bra \rho \ket_{+}-\bra \rho \ket_{-}={e\over g}b . \label{e29}
\eeq
Since the l.h.s. of eq.(\ref{e29}) is a complicated function of b, it
is solved graphically (see fig.(1)). It is seen that if the external
 magnetic field is
strong enough to satisfy $B^2 > {{m^2g}\over{\pi e^3}}$, at least one
non-zero solution exists which minimizes the effective potential. One such
solution, which is for the situation where $B^2 \simeq {{m^2g}\over{\pi e^3}}$,
is $b_0=\omega_c $. This, as has been discussed earlier in this section, leads
to an effective Zeeman interaction term ${{\omega_c }\over 2}\psi \dag \sigma_3
\psi $. Thus, we have the rather interesting result that a four fermi term in
 a strong external magnetic field can lead to an effective Zeeman interaction
for the fermions.

5. {\bf Relativistic Q.E.D. }

Having demonstrated the efficacy of our technique for the nonrelativistic
case, we now go on to discuss the relativistic scenario. Although this
is deemed unnecessary for most condensed matter oriented applications,
there are situations where the fermionic dispersion relation may be
taken to be approximately linear (e.g. in the neighbourhood of the Fermi
surface) and our subsequent calculations would find applicability in
precisely these cases. Furthermore, as has been discussed at length in the
introduction, the astrophysical and cosmological ramifications of this
problem, namely fermions interacting through a gauge field, Abelian or
otherwise, in a strong magnetic background, are manifold. The strong
magnetic field can induce fundamental changes not only in the vacuum structure
of such theories but also in the other sectors of the Fock space.

Admittedly, in this work, we have addressed the problem of only planar
fermions with an Abelian interaction. However, adding on the third
spatial dimension entails only very minor modifications, namely, the single
particle energies also come to depend quadratically on the z-component
of the momentum. This can be incorporated quite easily within our
computational framework as can the extension to non-Abelian interactions,
which actually bring in little more than notational complexity.

Having enumerated these motivations, we now proceed to the actual computation.
 The
generating functional is then given by
\beq
{\cal Z} = \int D\psi D\psib  e^{-\int d{\bf x} dt \psib \{ \Ds -m-\mu
\gamma^{0
} \} \psi } \equiv e^{-S_{eff}}
\eeq
where $\mu $ is the chemical potential and we use an euclidean metric with
signature $-1$.
The fermionic degrees of freedom can then be integrated out to yield the formal
expression of the effective action as
\beq
S_{eff} = -\mbox{Trlog}[ \Ds -m-\mu \gamma^{0} ]
\eeq
Define $\tilde{a}_{0} = a_{0} - \frac{i\mu }{e} $,
and $\tilde{\bf a } = {\bf a} $.
Then we can express the effective action as
\bqq
S_{eff} = -\mbox{Trlog} [ \tilde{\Ds } -m ]  \, ,
\eqq
with
\beq
 \tilde{D}_{\mu }= ip_{\mu } - ie \tilde{a}_{\mu } \equiv i\Pi _{\mu } .
\eeq
Note that the commutator $[\Pi_{1} ,\Pi_{2} ] = \frac{i}{l^{2}} $.

Upon functionally differentiating with respect to the gauge fields,
the corresponding current is obtained as
\beq
\ln J_{\mu} \rn  = \frac{ie}{2} \mbox{Tr} \ln x|\{ \gamma_{\mu } ,
\frac{1}{i\Ps -m}\} |x \rn
\eeq
The derivative expansion can now be performed in an identical manner to the
nonrelativistic case. The bra and ket are translated to the spacetime origin,
and the operators $\hx_{\mu } $ to $\hx_{\mu } +x_{\mu }$. We then formally
expand all functions of $\hx_{\mu}+x_{\mu}$ around $x_{\mu}$. Also, the unitary
transformations
U, V, and W on the current operator are identical to those in the
non-relativistic
case. Upon performing these transformations the current takes the form:

\beq
\langle J_{\mu }\rangle = \frac{ie}{2} Tr\langle 0|\{ \gamma_{\mu } ,\frac{1}{i
\Ps -m-\Os } \} |0\rangle
\eeq
with
\beq
O^{0} = ie\hx_{i} \partial^{i} a^{0} +\frac{ie}{2}\hx_{i} \hx_{j} \partial^{i}
 \partial^{j} a^{0} +ie\hx_{0} \hx_{i} \partial^{i} \partial^{0} a^{0} +\ldots
\eeq
and
\beq
O^{l} = \frac{ie}{2} \hx_{i} \partial^{i} a^{l} +\frac{ie}{3} \hx_{i} \hx_{j}
\partial^{i} \partial^{j} a^{l} +ie\hx_{0} \partial^{0} a^{l}+ie\hx_{0} \hx_{
i} \partial^{0} \partial^{i} a^{l}+\frac{ie}{2} \hx_{0}^{2} (\partial^{0} )^{2}
a^{l}
\eeq
Again, we can use perturbation theory to expand :
\beq
\frac{1}{i\Ps -m-\Os } = \frac{1}{i\Ps -m} +\frac{1}{i\Ps -m} \Os \frac{1}{i\Ps
 -m} +\ldots
\eeq
Also, the algebraic relations between $\gamma $ matrices (see appendix {\bf A})
 can
be used to express:
\beq
\frac{1}{i\Ps -m} = \frac{1}{\Pi^{2} -m^{2}+\frac{1}{l^2}\gamma^{0}} (i\Ps +
m)
\eeq
Defining
\bqq
S&=& \Pi^{2} -m^{2} \\ \nonumber
M_{0}& =& (\Pi^{2} -m^{2})^{2} - \frac{1}{l^{4}} \\ \nonumber
Q_{1}& =& \frac{S}{M_{0}} \\ \nonumber
Q_{2}& =& \frac{1}{l^{2}M_{0}}
\eqq
we obtain
\beq
\frac{1}{i\Ps -m}  = (Q_{1}+\gamma_{0} Q_{2})(i\Ps +m)
\eeq
and therefore
\bqq
\ln J_{\mu }(x) \rn &=&
 \frac{ie}{2} Tr\bra \{ \gamma_{\mu } ,(Q_{1}+\gamma_{0} Q_{2})
(i\Ps +m)  \\ \nonumber
&+& (Q_{1}+\gamma_{0} Q_{2})(i\Ps +m)\Os (Q_{1}+\gamma_{0} Q_{2})(i\Ps +m) \}
 \ket+\ldots \label{jm}
\eqq

In order to calculate expectation values in the state $\ket $, it will be
necessary to expand this state in the eigenbasis of the operator $\Pi^{2} $.
It is instructive to compute the eigenspectrum of $\Pi^{2} $ explicitly
since this calculation will also motivate the use of guiding center coordinates
{}.

Consider the eigenvalue equation
\beq
(\Pi^{2} -m^{2})\Psi  = \lambda \Psi
\eeq
i.e.
\beq
[ ( p_{0}+i\mu )^{2} + p_{1}^{2} + (p_{2}-eBx_{1})^{2} ] \Psi
 = -(\lambda +m^{2})\Psi
\eeq
One can solve this by separation of variables:
\beq
\Psi(x_{1},x_{2},x_{0})  = e^{-i\omega x^{0}}e^{-ieBXx^{2}}\Phi (x_{1})
\eeq
This leads to the single differential equation for $\Phi $:
\beq
[ -\partial_{1}^{2} + (eB)^{2}(x^{1}-X)^{2}]\Phi  = -[\lambda  +m^{2}+(\omega -
i\mu )^{2}]\Phi
\eeq
This is just the Schr\"{o}dinger equation for a harmonic oscillator in one
dimension.,
with eigenvalue spectrum $(2n+1)/l^{2}$. The full eigenvalue spectrum of the
operator $\Pi^{2} -m^{2}$ is therefore
\beq
\lambda_{n} = -\frac{(2n+1)}{l^{2}} - m^{2} - (\omega -i\mu )^{2}
\eeq
The corresponding eigenstates may be labelled by the quantum numbers $n$,
$\omega $
and $X$. It follows that
\bqq
S|X, n, \omega \rn & =& \lambda_{n} |X, n, \omega \rn \\
M_{0}|X, n, \omega \rn & =& \Gamma_{n} |X, n, \omega \rn
\eqq
where  $\Gamma_{n} = \lambda_{n}^{2} -\frac{1}{l^{4}}$
Therefore,
\bqq
Q_{1}|X, n, \omega \rn &=& \frac{\lambda_{n}}{ \Gamma_{n} } |X, n, \omega \rn
 \\ \nonumber
Q_{2}|X, n, \omega \rn &=& \frac{1}{l^{2} \Gamma_{n}} |X, n, \omega \rn
\eqq
and the completeness relation takes the form
\beq
\sum_{n} \int \frac{d\omega }{2\pi } dX|X, n, \omega \rn \ln X, n, \omega | = 1
\eeq
We now define the operators
\bqq
\hat{X} &\equiv &-\hx_{1} -l^{2}\Pi_{2} \equiv -l^{2}p_{2} \\ \nonumber
 \hat{Y} &\equiv &-\hx_{2} +l^{2}\Pi_{1} \equiv -\hx_{2} +l^{2}p_{1}
\label{gcent}
\eqq
They have the properties
\bqq
\hX |X, n, \omega \rn &=& X|X, n, \omega \rn \\ \nonumber
[\hX ,\hY ] & =& il^{2} \\ \nonumber
\hX \ket &=& -l^{2}\Pi_{2} \ket \\ \nonumber
\hY \ket &=& l^{2}\Pi_{1} \ket
\eqq
As we shall later see, these properties give us a powerful method of computing
expectation values of quantities in the state $\ket $.

Consider eq.(\ref{jm}).  Since the trace operation is independent of the
ordering
of operators in the commutator, the expression for the current to first
 order in
$O$ is given by:
\beq
\ln J_{\mu } \rn  = ie \mbox{Tr} \bra \gm (Q_{1}+\gamma_{0} Q_{2})(i\Ps +m)
\Os (Q_{1}+\gamma_{0} Q_{2})(i\Ps +m) \ket
\eeq
This gives rise to 16 separate terms. The calculation is done in the following
manner: First, in each term, the trace is taken with respect to the $\gamma $
matrices.
This leaves products of $Q$'s ,$\Pi $'s and $O$ in each term. The dependence on
the $x_{\mu}$ in $O$ is removed through the use of properties eq.(\ref{gcent})
of the guiding center
coordinates, as will be shown in a specific example below. Then the state
$\ket $ is expanded in the eigenstates of $Q_{1}$. Remembering that $\Pi_{1,2}
 $ act as combinations of raising and lowering operators on these eigenstates,
 the expectation
value can be computed in a straightforward, albeit tedious, manner.

To illustrate, consider the term $-ie\bra \gamma_{0} Q_{1}\Ps \hat{O}
\gamma_{0}
Q_{2}\Ps \ket $ in $\ln J_{0} \rn $. Upon taking the trace with respect to the
$\gamma $
matrices, this splits into 6 separate terms:
\bqq
& & 2e\bra Q_{1}\Pi_{1} O^{0}Q_{2}\Pi_{2} \ket  - 2e\bra Q_{1}\Pi_{2}
O^{0}Q_{2}
\Pi_{1} \ket  - 2e\bra Q_{1}\Pi_{0} O^{1}Q_{2}\Pi_{2} \ket   \nonumber \\
& & + 2e\bra Q_{1}
\Pi_{0} O^{2}Q_{2}\Pi_{1} \ket  + 2e\bra Q_{1}\Pi_{1} O^{2}Q_{2}\Pi_{0} \ket
- 2e\bra Q_{1}\Pi_{2} O^{1}Q_{2}\Pi_{0} \ket  \nonumber \\
& & \label{30}
\eqq
We would like to obtain a contribution proportional to $f^{12}$ from these
terms as a demonstration. We note that contributions to $f^{12}$ are obtained
from $O_1$ and $O_2$:
\[ O_1 \rightarrow \frac{ie}{2} \bfx_2 f^{12}
 \hspace{3cm}
O_2 \rightarrow - \frac{ie}{2} \bfx_1 f^{12}
\]
Thus the terms from eq.(\ref{30}) which contribute to $f^{12}$ are:
\[
ie^2 [ \ln 0| Q_1 \bfx_2 Q_2 \bfp_2 \hat{\Pi}_0|0 \rn+
 \ln 0| Q_1 \bfp_2 \bfx_2 Q_2 \hat{\Pi}_0|0 \rn
+ \ln 0| Q_1 \bfx_1 Q_2 \bfp_1 \hat{\Pi}_0|0 \rn
+ \ln 0| Q_1 \bfp_1 \bfx_1 Q_2 \hat{\Pi}_0|0 \rn ] f^{12}
\]
Using eq.(\ref{gcent}) and the commutators in  appendix {\bf B} we get
\[ -\frac{4e^2}{l^2} [ \ln 0| \frac{S}{M_0} \{ S, \bfp_1 \}
\frac{1}{M_0^2}\bfp_1 \hat{\Pi}_0|0 \rn
-\ln 0|\frac{S}{M_0^2}\bfp_1 \frac{1}{M_0}
\{ S, \bfp_1 \} \hat{\Pi}_0|0 \rn -\ln 0| \bfp_1 \frac{1}{M_0^2} \bfp_1
\hat{\Pi}_0|0 \rn  ]f^{12} .
\]
Now there is another term in $\ln J_0 \rn$  with $ Q_1 \rightarrow Q_2$ . This
contributes to
$f^{12}$ in the form:
\[ -\frac{4e^2}{l^2} [ \ln 0| \frac{1}{M_0} \{ S, \bfp_1 \}
\frac{S}{M_0^2}\bfp_1 \hat{\Pi}_0|0 \rn
+\ln 0|\frac{1}{M_0^2}\bfp_1 \frac{S}{M_0}
\{ S, \bfp_1 \} \hat{\Pi}_0|0 \rn -\ln 0| \frac{1}{M_0} \bfp_1^2 \frac{1}{M_0}
\hat{\Pi}_0|0 \rn
]f^{12}.
\]
Consider the simplest term:
\bqq
& &\frac{4e^2 }{l^2} f^{12}
\ln 0|\bfp_1 \frac{1}{M_0^2} \bfp_1 \hat{\Pi}_0|0 \rn = \\ \nonumber
& &\frac{4e^2 }{l^2} f^{12}  \sum_{n,n_1,n_2} \int \frac{dw}{2 \pi}(w-i\mu)
\int
dX \frac{ \ln 0|n,X,w \rn \ln n|\bfp_1 |n_1 \rn \ln n_1 |\bfp_1 |n_2 \rn \ln
n_2,X,w|0 \rn }
{\G^2_{n_1}}
\eqq
The integration over $X$ now gives $\rho_0$ so we get
\bqq
 \frac{2e^2}{\pi l^4} f^{12} \sum_{n_1,n}& \int & \frac{dw}{2 \pi}( w-i\mu)
 \frac{ \ln n|\bfp_1 |n_1 \rn \ln n_1 |\bfp_1 |n \rn }
{\G_{n_1}^2} \\ \nonumber
&=& \frac{e^2}{\pi l^6} f^{12} \sum_{n} \int \frac{dw}{2 \pi}(w-i\mu) \left[
\frac{n+1}
{\G^2_{n+1}} +\frac{n}{\G^2_{n-1}}\right] \\ \nonumber
&=&\frac{e^2}{\pi l^6} f^{12} \sum_{n} \int \frac{dw}{2 \pi}(w-i\mu)(n+1)
\left[ \frac{1}
{\G^2_{n+1}} +\frac{1}{\G^2_{n}} \right]
\eqq
simplifying the other terms in the same way we get
\[
- \frac{2 e^2}{\pi l^6} f^{12} \sum_{n=0}^{\infty} \int \frac{dw}{2 \pi} (w-i
\mu)
\frac{(n+1)(\lambda_n+\lambda_{n+1})}{\G_n \G_{n+1}} \left(
\frac{\lambda_n}{\G_n}
+\frac{\lambda_{n+1}}{\G_{n+1}} \right)
\]
The coefficient of the term proportional to $f^{12}$ in $\ln j_0 \rn$ also
receives
 contributions
proportional to $m$. These are calculated to be
\bqq
\frac{2ie^2 }{\pi l^2 w_c} f^{12} \sum_{n=0}^{\infty} \int \frac{dw}{2 \pi}
\left[ (n+1)\frac{( \L_n+\L_{n+1})}{\G_n \G_{n+1}} \left( \frac{\L_n (\L_{n+1}-
\frac{1}{l^2})}{\G_n}+\frac{\L_{n+1} (\L_{n}-\frac{1}{l^2})}{\G_{n+1}} \right)-
(2n+1)\frac{\L_n}{\G_n^2} \right]
\eqq
A further simplification is in order in the above expressions. Let us define
\[ d_n \equiv -\frac{2n}{l^2}-m^2-(w-i \mu)^2 \, . \]
  Thus
\beq
\frac{1}{\G_n} = \frac{1}{\L_n^2-\frac{1}{l^4}}=\frac{l^2}{2} \left(
\frac{1}{d_{n+1}}
-\frac{1}{d_n} \right)
\eeq
using this technique, any denominator containing arbitrary products of the
$\G$s can
be simplified to powers of a given $d$. Further, the numerators can trivially
rewritten in terms of $d$ since  $\L_n=d_n-\frac{1}{l^2}$.

The final expression  for $\ln J_0 \rn$ and $\ln J_1 \rn$ can be expressed
entirely
in terms of $w$ , $\mu$ and $d_n$.
It is quite straightforward to show, as in the nonrelativistic case, that the
terms that potentially violate gauge invariance vanish identically.
Compiling all the terms together we get:
\bqq
\ln J_0 \rn &=&  C_0 f^{12} +C_1 \partial^0 \partial^1 a^1 + C_2
 (\partial^1)^2 a^{0} +\ldots
\\ \nonumber
\ln J_1 \rn &=&  D_0 \partial^2 f^{12} +D_1 \partial^0 \partial^1 a^0 + D_2
 (\partial^0)^2 a^{1} +\ldots
\eqq
where $C_0 $ , $C_i$ and $D_0$ , $D_i$ are constants to be specified later.
Correspondingly,
\bqq
\ln \Pi_{00}(x,y) \rn &=& C_2 (\partial^1)^2 \delta (x-y)+C_2 (\partial^2)^2
\delta (x-y)+\ldots
\\ \nonumber
\ln \Pi_{01}(x,y) \rn &=& -C_0 \partial^2 \delta (x-y)+C_1 \partial^0
 \partial^1 \delta (x-y)+\ldots
\\ \nonumber
\ln \Pi_{11}(x,y) \rn &=& -C_0 (\partial^2)^2 \delta (x-y)+D_2 (\partial^0)^2
 \delta (x-y)+\ldots
\\ \nonumber
\ln \Pi_{12}(x,y) \rn &=& -D_0 \partial^2 \partial ^1 \delta (x-y)+\ldots
\\ \nonumber
\ln \Pi_{02}(x,y) \rn &=& C_0 \partial^1 \delta (x-y)+C_1 \partial^0
\partial^2 \delta (x-y)+\ldots
\\ \nonumber
\ln \Pi_{22}(x,y) \rn &=& -D_0 (\partial^1)^2 \delta (x-y)+D_2
(\partial^0)^2 \delta (x-y)+\ldots
\eqq
Gauge invariance forces $C_1=-C_2=D_1=-D_2 \equiv C$  so that the effective
action that
one gets after collecting all the terms and functionally integrating $\ln
\Pi_{\mu \nu} \rn$
is
\beq
S_{eff}=\frac{C}{2} \int d^3 x \, {\cal E}^2+ \frac{D}{2} \int d^3 x \, {\cal
B}^2
+\frac{C_0}{2} \int d^3 x \, a^{0} {\cal B}+ \ldots
\label{relef}
\eeq
where
\beq
D_0 \equiv D=-\frac{e^2}{12 \pi} \sum_{n=0}^{\infty} \left[ \frac{m}{2 w_c}+
7n-\frac{5}{2} \right] \frac{1}{\sqrt{m^2+2neB}}
\Theta (\sqrt{m^2+2neB}-| \mu | ) \, \, ,
\eeq
 \[ C_0=\frac{e^2}{2 \pi} \sum_n \int dw \,
 \frac{w-i\mu}{d_n}=-\frac{ie^2}{2} \frac{\mu}{| \mu |} \sum_n \Theta (| \mu |
-\sqrt{2meB+m^2})
\] and
\bqq
C&=&\frac{e^2 m}{32 \pi w_c} \sum_{n=0}^{\infty} \frac{(10-13n)}{
\sqrt{m^2+2neB}}\Theta (\sqrt{m^2+2neB}-| \mu | )
\\ \nonumber
&-& \frac{e^2}{16 \pi} \sum_{n=0}^{\infty}  \frac{n(10-13n)}{
\sqrt{m^2+2neB}}\Theta (\sqrt{m^2+2neB}-| \mu | )
\eqq
One sees that even in the relativistic case the coefficient of the euclidean
${\cal E}^2$  vanishes for strong external magnetic field. The explicit
dependence on the chemical potential $\mu$ can be removed by specifying
the number of filled Landau levels. Physical quantities can be extracted
from this action as in the nonrelativistic case.

{\bf Discussions}

In this paper we have calculated the effective action of nonrelativistic
and relativistic 2+1 dimensional QED in a strong external magnetic
field using the inhomogeneity expansion method. We have explicitly addressed
only the
case of  the parity odd model i.e. one fermion flavour , but it is
straightforward to use our method in the parity even case by considering
two fermion flavours with positive and negative mass \cite{deser}. For
the nonrelativistic case we obtain a Chern-Simons like term in
addition to the renormalization of the coefficients of ${\cal E}^2$  and
${\cal B}^2$. In the relativistic case a true Chern-Simons term is generated
and we calculate its coefficient as well as the leading term in the
coefficients of  ${\cal E}^2$ and ${\cal B}^2 $.
For very stong external magnetic fields ($l \ll \frac{1}{m^2}$) we see that
the coefficient of the euclidean ${\cal E}^2$ term vanishes both in the
relativistic and nonrelativistic case. This suggests that a dimensional
reduction is indeed taking place \cite{iso}. In this context, it would
be interesting to study the effect of such a reduction in the case of
Nambu Jona-Lasinio models in general and the 2+1 dimensional Thirring
model in particular \cite{rg}.

We have extracted just a few of the many possible physical quantities
that can be obtained from these effective actions {\it viz} the dielectric
constant and magnetic permeability. Considering a finite temperature
variant \cite{saki}, we can further discuss the physics of plasmons
in a magnetic field background. A particularly intriguing direction is to
adapt these techniques to astrophysical and cosmological applications.

{\bf Acknowledgements}

We wish to thank Jonathan Simon for interesting discussions. This work is
partially supported by the N.S.F.

{\bf Appendix A.  Dirac algebra and Trace relations}

In the relativistic calculation, the metric we use is $g_{\mu \nu } = $
diag$(-1,-1,-1)$.
One has to be careful,
therefore, in raising and lowering indices. Every raising or lowering of an
index picks up a negative sign. The canonical commutation relation takes the
form
\beq
[ \hx_{\mu } ,\hp_{\nu } ]  = ig_{\mu \nu }
\eeq
from which commutation relations involving $\Pi $'s follow directly.

The algebra of Dirac matrices in 2+1 dimensions, with a negative definite
Euclidean metric, albeit straightforward, is not generally familiar. Here we
quote the main relations which are used in the body of the paper.

To start with, the Dirac matrices are defined as $\gamma_{1,2} =\sigma_{1,2} $,
 $\gamma_{0} =\sigma_{3} $, where the $\sigma $'s are the usual Pauli matrices.
(note that $\gamma^{i} =-\sigma_{i} $, because the metric has negative definite
 signature.)
This leads to the relations
\beq
\gamma_{\mu } \gamma_{\nu } = -g_{\mu \nu } + i\epsilon_{\mu \nu \rho } \gamma_
{\rho }
\eeq
Hence
\beq
\{ \gamma_{\mu } ,\gamma_{\nu } \} = -2g_{\mu \nu}.
\eeq
The trace relations are obtained as
\bqq
\mbox{Tr}\gm \gn &=& -2g_{\mu \nu } \\ \nonumber
\mbox{Tr}\gm \gn \gamma_{\lambda }& =& 2i\epsilon_{\mu \nu \lambda }
 \\ \nonumber
\mbox{Tr}\gm \gn \ga \gb & =& 2(g_{\mu \nu }
g_{\alpha \beta } - g_{\mu \alpha }g_{\nu \beta } +
 g_{\mu \beta }g_{\nu \alpha }
) \\ \nonumber
\mbox{Tr}\gm \gn \ga \gb \gd & =& -2i ( g_{\mu \nu }\epsilon_{\alpha
\beta \delta } + g_{\alpha \beta }\epsilon_{\mu \nu \delta } +
 g_{\beta \delta }\epsilon_{\mu \nu \alpha } - g_{\delta \alpha }
\epsilon_{\mu \nu \beta } )
\\ \nonumber
\mbox{Tr}\gm \gn \ga \gb \gr \gs &=& -2 ( g_{\mu \nu }g_{\alpha
 \beta }g_{\rho \sigma}- g_{\mu \alpha }g_{\nu \beta }g_{\rho \sigma }
 + g_{\mu \beta }g_{\nu \alpha }
g_{\rho \sigma } - g_{\alpha \beta }g_{\mu \rho }g_{\nu \sigma } \\ \nonumber
& &+ g_{\alpha
\beta }g_{\mu \sigma }g_{\nu \rho } - g_{\mu \nu }g_{\alpha \rho }g_{\beta
\sigma }
+ g_{\mu \nu }g_{\alpha \sigma }g_{\beta \rho } - \epsilon_{\mu \nu \rho }
\epsilon_{\alpha \beta \sigma } + \epsilon_{\mu \nu \sigma } \epsilon_{\alpha
\beta \rho } )
\eqq

and so on. The trace of an odd number of $\gamma $ matrices does not vanish as
it does in 3+1 dimensions.

{\bf Appendix B. Some useful relations}

First, for the non-relativistic case
\bqq
[\bfp_2, \frac{1}{H_0}]&=&\frac{i}{ml^2}\frac{1}{H_0} \bfp_1 \frac{1}{H_0}
\\ \nonumber
[\bfp_1,\frac{1}{H_0}]&=&-\frac{i}{ml^2} \frac{1}{H_0} \bfp_2 \frac{1}{H_0}
\\ \nonumber
[\hat{\tau},\frac{1}{H_0}]&=& \frac{1}{H_0^2}
\eqq
For the relativistic case
\bqq
[ \hx_0,Q_1]&=&-4iS\frac{1}{M_0}S \hat{\Pi}_0\frac{1}{M_0}+2i\hat{\Pi}_0
\frac{1}{M_0} \\ \nonumber
[ \hx_0, Q_2]&=& -\frac{4i}{l^2} \frac{1}{M_0} S \hat{\Pi}_0 \frac{1}{M_0}
\\ \nonumber
[\bfp_1,Q_2]&=&-\frac{2i}{l^4} frac{1}{M_0} \{ S, \bfp_2 \} \frac{1}{M_0}
\\ \nonumber
[\bfp_1,Q_1]&=&-\frac{2i}{l^2} \frac{S}{M_0} \{ S, \bfp_2 \} \frac{1}{M_0}+
\frac{2i}{l^2} \bfp_2 \frac{1}{M_0}
\\ \nonumber
[ \bfp_2, Q_1]&=& \frac{2i}{l^2} \left[ \frac{1}{M_0} \{ S, \bfp_1 \}
\frac{1}{M_0}- \bfp_1 \frac{1}{M_0} \right]
\\ \nonumber
[ \bfp_2, Q_2]&=& \frac{2i}{l^4} \frac{1}{M_0} \{ S, \bfp_1 \} \frac{1}{M_0}
\eqq


\newpage

\begin{thebibliography}{99}
\bibitem{Shapiro} S. L. Shapiro and S. A. Teukolsky, \lq \lq {\it Black Holes,
 White Dwarfs and Neutron Stars, The Physics of Compact Objects}'',
(Wiley, N.Y.) (1983).
\bibitem{Narayan} R. Narayan, P.  Paczynski and T. Piran, {\it  Ap. J. Lett.},
 395, {\bf L}83 (1992).
\bibitem{Ambjorn} J. Ambjorn and P. Olesen, {\it Nucl. Phys.} {\bf B }315,
606, (1989),  and references therein.
\bibitem{Skal} V. V. Skalozub, {\it Sov. J. Nucl. Phys.} {\bf 45}, 1058, (1987)
and referenc es therein.
\bibitem{Feldman} A. Erdas and G. Feldman, {\it  Nucl. Phys.} {\bf B }343,
(1990), 59 7.
 \bibitem{Damgaard} P.H. Damgaard and D. Espriu, {\it Phys. Lett.} {\bf B }246,
 442, (1991).
\bibitem{Krive} I.V. Krive and S. A. Naftulin, {\it  Phys. rev.} {\bf D }46,
 2737, (1992).
\bibitem{Klimenko} K.G. Klimenko, {\it Z. Phys.} {\bf C }54, (1992), 323.
\bibitem{Pani} P. K.  Panigrahi, R.  Ray and B. Sakita, {\it Phys. Rev.} {\bf B
}42,
 4036, (1990).
\bibitem{Schakel} A.M.J. Schakel,  \lq \lq On Broken Symmetries in Fermi
 Systems'', (1989) and references therein.
\bibitem{Adrian} A. Neagu and A.M.J. Schakel,  {\it Phys. Rev.}
 {\bf D }48, 1785, (1993).
\bibitem{iso} S.  Iso, D. Karabali and B. Sakita {\it Phys. Lett.}
 {\bf B} 296, 143, (1992); {\it Nuc. Phys. } {\bf B} 388, 700 , (1992)
\bibitem{rg} R. Ray and G. Gat, work in progress.
\bibitem{saki} S. Sakhi, \lq \lq {\it Effective action of a 2+1 dimensional
system of nonrelativistic fermions in the presence of a uniform magnetic
field: dissipation effects}" UdeM-LPN-TH-179.
\bibitem{deser} S. Deser, R. Jackiw and S. Templeton {\it Ann. Phys.} 140,
372, (1982).
\end{thebibliography}
\end{document}